# Electron Accumulation and Emergent Magnetism in LaMnO$_3$/SrTiO$_3$ Heterostructures

Zuhuang Chen[1,2], Zhanghui Chen[2], Z. Q. Liu[3*], M. E. Holtz[4], C. J. Li[5,6], X. Renshaw Wang[7], W. M. Lü[8], M. Motapothula[6], L. S. Fan[9], J. A. Turcaud,[1] L. R. Dedon,[1] C. Frederick[10], R. J. Xu,[1] R. Gao,[1] A. T. N'Diaye[11], E. Arenholz[11], J. A. Mundy[1,2], T. Venkatesan[5,6], D. A. Muller[4], L.-W. Wang[2], J. Liu[10*], and L. W. Martin[1,2*]

[1] Department of Materials Science and Engineering, University of California, Berkeley, California 94720, USA

[2] Materials Science Division, Lawrence Berkeley National Laboratory, Berkeley, California 94720, USA

[3] School of Materials Science and Engineering, Beihang University, Beijing 100191, China

[4] School of Applied and Engineering Physics, Cornell University, Ithaca, New York 14853, USA

[5] Department of Materials Science and Engineering, National University of Singapore, Singapore 117575, Singapore

[6] NUSNNI-Nanocore, National University of Singapore, Singapore 117411, Singapore

[7] School of Physical and Mathematical Sciences & School of Electrical and Electronic Engineering, Nanyang Technological University, Singapore 637371

[8] Condensed Matter Science and Technology Institute, School of Science, Harbin Institute of Technology, Harbin 150081, People's Republic of China

[9] Center for Nanophase Materials Sciences, Oak Ridge National Laboratory, Oak Ridge, Tennessee 37831, USA

[10] Department of Physics and Astronomy, University of Tennessee, Knoxville, Tennessee 37996, USA

[11] Advanced Light Source, Lawrence Berkeley National Laboratory, Berkeley, California 94720, USA

*Email: zhiqi@buaa.edu.cn; jianliu@utk.edu; lwmartin@berkeley.edu

**Abstract:**

Emergent phenomena at polar-nonpolar oxide interfaces have been studied intensely in pursuit of next-generation oxide electronics and spintronics. Here we report the disentanglement of critical thicknesses for electron reconstruction and the emergence of ferromagnetism in polar-mismatched $LaMnO_3/SrTiO_3$ (001) heterostructures. Using a combination of element-specific X-ray absorption spectroscopy and dichroism, and first-principles calculations, interfacial electron accumulation and ferromagnetism have been observed *within* the polar, antiferromagnetic insulator $LaMnO_3$. Our results show that the critical thickness for the onset of electron accumulation is as thin as 2 unit cells (UC), significantly thinner than the observed critical thickness for ferromagnetism of 5 UC. The absence of ferromagnetism below 5 UC is likely induced by electron over-accumulation. In turn, by controlling the doping of the $LaMnO_3$, we are able to neutralize the excessive electrons from the polar mismatch in ultrathin $LaMnO_3$ films and thus enable ferromagnetism in films as thin as 3 UC, extending the limits of our ability to synthesize and tailor emergent phenomena at interfaces and demonstrating manipulation of the electronic and magnetic structures of materials at the shortest length scales.

The observation of emergent interfacial electronic reconstruction and, in turn, new electronic phases at polar/non-polar oxide interfaces have launched numerous studies on the fundamental mechanisms for and potential uses of these exotic properties [1-3]. The nature of such electronic reconstruction (used here in the broad sense as any change in valence that results from the interfacial polar mismatch) depends on many factors such as the interfacial band alignment and the conductivity of the materials that can alter the boundary conditions and thus the degree of charge compensation [4]. To date, the most widely-studied system in this regard is the two-dimensional electron gas in $LaAlO_3/SrTiO_3$ [5]. $LaAlO_3$ is an insulator with a band gap of 5.6 eV which consists of alternating charged $(LaO)^+$ and $(AlO_2)^-$ layers along the [001]. $SrTiO_3$ is also an insulator, with a band gap of 3.2 eV, which has neutral $(SrO)^0$ and $(TiO_2)^0$ layers alternating along the [001]. When these materials are brought together, a so-called *polarization catastrophe* is believed to occur at the (001) interface due to a potential build-up in the polar $LaAlO_3$ which drives charge transfer from the $LaAlO_3$ valence band to the $SrTiO_3$ conduction band at a $LaAlO_3$ critical thickness of 4 unit cells (UC) [6-8]. As a result, there is electron-doping into the $SrTiO_3$ near the interface, which induces a reduction of $Ti^{4+}$ towards $Ti^{3+}$ [9] and the onset of conductivity [2]. Meanwhile, there have been reports of other exotic interfacial phenomena, such as orbital reconstruction [10], ferromagnetism [11], and superconductivity [12], and researchers have also explored the role that structural imperfections can play in the evolution of effects at these interfaces [13-17].

Since the initial observation of emergent phenomena at the $LaAlO_3/SrTiO_3$ heterointerface, a number of additional systems have been found to exhibit similar potential build-up induced electronic reconstruction and interfacial phenomena (*e.g.*, $LaTiO_3/SrTiO_3$ [18], $LaVO_3/SrTiO_3$ [19], $NdTiO_3/SrTiO_3$ [20], etc.). Studies of other related polar/non-polar interfaces, namely

LaCrO$_3$/SrTiO$_3$ [21] and LaMnO$_3$/SrTiO$_3$ [22], however, have suggested that this rule is not universal since no critical-thickness for a metal-to-insulator transition was found in either system. At first glance this is surprising since, for example, LaMnO$_3$ is constructed from alternating (LaO)$^+$ and (MnO$_2$)$^-$ layers and thus electronic reconstruction akin to that in LaAlO$_3$/SrTiO$_3$ should be expected. Initially, the lack of metallic conductivity was enough to discourage researchers from further study, but it does not necessarily mean that the interfacial polar mismatch does not have some influence on the charge or spin degrees of freedom. In fact, recent work on the LaMnO$_3$/SrTiO$_3$ system has proposed that charge transfer may occur within LaMnO$_3$ as a result of a similar potential build-up which produces electron doping near the interface and hole doping near the surface and subsequently gives rise to ferromagnetism in LaMnO$_3$ [23, 24]. Direct evidence for the electron reconstruction and the nature of magnetism has not, however, been reported, precluding further understanding and control of the functional properties.

In this work, we apply X-ray absorption spectroscopy (XAS) and magnetic circular dichroism (XMCD), and first-principles calculations to investigate the effect of polar mismatch on the electronic and magnetic structure of the polar, antiferromagnetic insulator LaMnO$_3$ grown on nonpolar SrTiO$_3$. XAS reveals a significant change of the Mn valence state in LaMnO$_3$ even in 2-UC-thick films, but no change in Ti valence in the SrTiO$_3$. XMCD, on the other hand, reveals ferromagnetism with a critical thickness of 5 UC. The decoupling of the charge and magnetic critical thicknesses and the absence of hole doping near the film surface suggest that the charge transfer picture [23, 24] may not be directly applicable or is an oversimplification. Theoretical studies show that the potential build-up in the LaMnO$_3$ is 0.177 V/Å and this, together with the small band gap of LaMnO$_3$, results in a critical thickness for the built-in potential to collapse at only 2 UC. In turn, we identify a chemical route – whereby small changes in film stoichiometry – can be used to effectively control the doping level and modulate the influence of the polar mismatch so as to induce ferromagnetism in films down to just 3 UC.

LaMnO$_3$ films were grown on TiO$_2$-terminated SrTiO$_3$ (001) substrates by reflection high-energy electron diffraction (RHEED)-assisted pulsed-laser deposition [25]. For all films, RHEED intensity oscillations were persistent throughout the growth indicating a layer-by-layer growth mode [25]. Further characterization of the heterostructures reveal high-quality LaMnO$_3$ films with atomically-smooth surfaces and high crystalline quality, and interfaces that are free of dislocations [25]. To probe the potential for electron reconstruction and associated ferromagnetic order, we have performed soft XAS and XMCD studies in a grazing incidence, total electron yield (TEY) geometry on numerous LaMnO$_3$/SrTiO$_3$ heterostructures (inset, Fig. 1a) [25]. Compared to previous local magnetometry studies using a scanning superconducting quantum interference device (SQUID) [23, 24], the current approach is surface sensitive and provides a direct measurement of the valence state of the various elements in the films and substrates *via* XAS and enables the detection of subtle (~ 0.005 $\mu_B$/atom), element-specific magnetic moments (thus excluding magnetic impurities) *via* XMCD [9]. XMCD measurements at the Mn $L_{2,3}$ edges at 25 K reveal strong dichroism (Fig. 1a), indicative of a net ferromagnetic moment, arising from Mn in the films. XMCD of ~28% and ~26% is observed for the 9 and 12 UC thick LaMnO$_3$ films, respectively. To give this some context, the corresponding SQUID magnetometry measurement on the 9 UC thick LaMnO$_3$/SrTiO$_3$ heterostructure revealed a magnetic moment of ~400 emu/cc at 25 K [25]. The XMCD is reduced to ~22% for the 6 UC films and finally to ~11% for the 5 UC films. There are a number of important observations to be noted here. First, due to the high sensitivity of the XMCD, we are able to detect the small magnetic moment of the 5 UC thick films, something not accomplished in previous studies [23]. Second, the large XMCD in the LaMnO$_3$ films as thin as 9 UC is comparable to that observed in hole-doped La$_{0.67}$Sr$_{0.33}$MnO$_3$ [40].

These thickness-dependent XMCD results confirm the existence of intrinsic magnetization in the LaMnO$_3$ films, in line with previous magnetometry studies [23, 24]. To understand the nature of the magnetism, we first performed XAS measurements at both the Ti- and Mn-$L_{2,3}$ edges as a function of LaMnO$_3$ thickness to probe the corresponding evolution of the valence states. Regardless of the LaMnO$_3$ thickness, the Ti-$L_{2,3}$ edges are unchanged and match that of a SrTiO$_3$ substrate reference (Fig. 1b). In other words, there is no indication of electron transfer to the SrTiO$_3$ and the bulk-like character of Ti$^{4+}$ is maintained at the LaMnO$_3$/SrTiO$_3$ heterointerface. In contrast, XAS studies at the Mn-$L_{2,3}$ edges reveal marked variation with film thickness (Fig. 1c). To understand the evolution of the Mn valence state, we provide spectra for bulk SrMnO$_3$ (*i.e.*, Mn$^{4+}$ reference), LaMnO$_3$ (*i.e.*, Mn$^{3+}$ reference), and MnO (*i.e.*, Mn$^{2+}$ reference) for comparison [41, 42]. Starting with relatively thick LaMnO$_3$ films (*e.g.*, 12 UC), the absorption spectra are observed to be similar to that of bulk LaMnO$_3$, indicating a predominant Mn$^{3+}$ valence state. Note that the absorption peak has a slightly lower energy position than that of bulk LaMnO$_3$, indicating that the films are somewhat electron doped. Upon reducing the LaMnO$_3$ thickness (*e.g.*, 9 UC) a shift of the main spectral feature near 641.4 eV to lower energies is observed, indicating a lower average valence state. Further reducing the LaMnO$_3$ thickness (*e.g.*, 6, 5, and 3 UC) results in further shifts of the feature. Moreover, a strong feature near 640 eV appears which corresponds to the main absorption peak of Mn$^{2+}$. Note that the suppression of the fine feature at ~638 eV in our ultrathin LaMnO$_3$ films, as compared to the MnO reference spectrum, is likely due to different local environments for the two structures [43]. Similar studies of 2 UC thick heterostructures likewise reveal a strong feature from Mn$^{2+}$ [25]; these data are not shown on the same graph since sample charging distorts the background of the spectra. Our XAS studies clearly show that the average Mn valence state within the thin samples is reduced, and further suggest that electrons are accumulated in the LaMnO$_3$ layer near the interface starting at a critical thickness of 2 UC. This is in contrast to our above XMCD studies and the previous local magnetometry studies [23, 24], which found that the emergent ferromagnetism is not measureable until a thickness 5-6 UC. That is, there is a decoupling of the electron accumulation and ferromagnetic critical thicknesses.

Furthermore, our studies are completed in TEY geometry which probes the topmost 3-5 nm of thin-film samples such that the contribution of the deeper layers to the TEY intensity decays exponentially with the distance from the film surface. Therefore, with increasing LaMnO$_3$ layer thickness the contribution of the Mn at the interface decreases accordingly. The observed decrease of Mn$^{2+}$ signatures in the Mn XAS spectra with increasing LaMnO$_3$ layer thickness suggests that the valence state is likely non-uniform along the surface normal. Thus, we find that the fraction of Mn$^{2+}$ is higher in the thinner films where the interface contributes more to the Mn XAS signal. This variation in valence state with film thickness is also evident in the evolution of the O $K$ edge [25]. It is possible that the bottom 3 UC, near the interfacial region, remains Mn$^{2+}$-rich for the 5 and 6 UC samples [44]. As the thickness increases beyond 6 UC, the thickness becomes comparable or larger than the probing depth and the contribution from the interfacial region to the spectrum is reduced. To compensate for the polar field, the excess electrons are likely to accumulate at the interface, creating an inhomogeneous charge profile across the film thickness. Such an observation is consistent with studies of La$_{1-x}$Sr$_x$MnO$_3$/SrTiO$_3$ (001) heterostructures wherein excess electrons are found to reside near the interfaces [44].

The presence of this electron accumulation due to the polar mismatch is further supported by similar XAS studies of LaMnO$_3$/NdGaO$_3$ (110)$_O$ heterostructures (subscript O denotes orthorhombic index). The NdGaO$_3$ substrates were treated to produce GaO$_2$-terminated surfaces which are polar in nature [45], consequently, there should be *no* polar mismatch at the LaMnO$_3$/NdGaO$_3$ heterointerface. XAS spectra at the Mn-$L_{2,3}$ edges are characteristic of bulk-like Mn$^{3+}$ with no Mn$^{2+}$ features (even for 3 UC thick films), and there is no thickness dependence of the XAS spectra (Fig. 1d); thus no excess charge is observed in the LaMnO$_3$/NdGaO$_3$ heterostructures. Ultimately, these studies indicate that electrons accumulate in the LaMnO$_3$ near the interfaces to alleviate the *polar catastrophe* in the LaMnO$_3$/SrTiO$_3$ heterostructures at film thicknesses as small as 2 UC.

To understand the origin of the electron accumulation, the band offset at the LaMnO$_3$/SrTiO$_3$ interface and the potential build-up in the LaMnO$_3$ were computed for symmetric (LaMnO$_3$)$_m$/(SrTiO$_3$)$_n$/(LaMnO$_3$)$_m$ ($n = 4.5$, $m = 2, 4, 6$ UC) structures (Fig. 2a, $m = 4$ UC) using first-principles density functional theory (DFT) [25]. The resulting in-plane-averaged (oscillating blue line) and macroscopically-averaged (red line) electrostatic potentials reveal a potential build-up in the LaMnO$_3$ layers due to the alternating charged atomic layers along the surface normal (Fig. 2b, $m = 4$). An average slope for the potential evolution was extracted for heterostructures with $m = 2, 4$ (Fig. 2c) from which the average internal field is determined to be ~0.177 V/Å. This internal field is smaller than that reported for the LaAlO$_3$/SrTiO$_3$ system (0.24 V/Å) [7] and the difference could be due to the different screening effects of LaAlO$_3$ and LaMnO$_3$. The calculated potential build-up is based on an idealized abrupt interface without defects and adsorbates. Such a potential build-up, however, has been difficult to detect experimentally (see Figs. 35-37, Ref. [15]) and is still under debate. Additionally, by evaluating the average potential in bulk and multi-layer heterostructures, the valence band offset between the LaMnO$_3$ and SrTiO$_3$ was extracted to be ~0.6 eV [25]. This, combined with the large difference in band gap between LaMnO$_3$ (1.3 eV) and SrTiO$_3$ (3.2 eV) [25], results in a straddling band configuration at the LaMnO$_3$/SrTiO$_3$ interface (Fig. 2d). Based on these two facts, electron accumulation is expected to occur when the built-in potential in the LaMnO$_3$ layer surpasses the LaMnO$_3$ band gap, which is calculated to happen at a critical thickness of ~7.3 Å (just less than 2 UC of LaMnO$_3$). As a result, the system will become unstable and must find a way to compensate the potential. Because the potential difference is positive from the interface to the surface, the screening or compensation effect will always tend to accumulate negative charges near the interface. This naturally explains the reduction of the Mn$^{3+}$ towards Mn$^{2+}$ in the very thin limit and near the interfaces for thicker samples. This is different from the case of LaAlO$_3$/SrTiO$_3$, where the LaAlO$_3$ band gap straddles the SrTiO$_3$ bands, thus the negative charges stay in the SrTiO$_3$ near the interface and reduce the Ti$^{4+}$ towards Ti$^{3+}$ [7-9].

Intriguingly, no indication of Mn$^{4+}$ was detected at/near the LaMnO$_3$ film surfaces, implying that the previous proposed charge transfer model [23, 24] could be oversimplified or may not be directly applicable to this system. In turn, any number of additional surface structural or chemical reconstructions could possibly occur to adequately compensate for the electron accumulation [46-49]. Furthermore, oxygen vacancies, in particular, have been considered as a means to compensate the polar field [48-50]. In these LaMnO$_3$/SrTiO$_3$ heterostructures, there is no evidence for the reduction of and oxygen vacancy formation in the SrTiO$_3$ since there was no signature of Ti$^{3+}$ detected via XAS near the interface and the heterostructures show insulating behavior [25]. Furthermore, there is no evidence for a large concentration of oxygen vacancies in either the film or substrate as a result of the growth process. This is supported by the lack of evidence for Mn$^{2+}$ in LaMnO$_3$/NdGaO$_3$ heterostructures which, despite being fabricated at the same conditions, show no evidence of a change in the valence state. Both of these observations are consistent with the high oxygen growth pressure (10$^{-2}$ mbar) [25]. In turn, the emergence of Mn$^{2+}$ in the LaMnO$_3$/SrTiO$_3$ heterostructures is attributed to the polar discontinuity at the heterointerface regardless of the specific mechanism responsible for the electron accumulation which could be electronic, structural, and/or chemical in nature [46-51]. Oxygen vacancies (not necessarily induced by deposition) have been proposed to be an intrinsic compensation mechanism for polar/nonpolar oxide heterostructures, which would lead to electron accumulation at the interface [48, 49, 51].

Bulk, stoichiometric LaMnO$_3$ with Mn$^{3+}$ is an *A*-type antiferromagnetic insulator [52]. Despite a possible electron-hole asymmetry in the manganite phase diagram similar to cuprates [53], the addition of either extra holes or electrons via chemical doping could lead to ferromagnetism as a result of Mn$^{3+}$-O$^{2-}$-Mn$^{4+}$ or Mn$^{3+}$-O$^{2-}$-Mn$^{2+}$ double exchange, respectively [52, 54-57]. It is known that extrinsic defects, such as cation-deficiency (*e.g.*, La-deficiency) or oxygen-excess, could give rise to ferromagnetism in LaMnO$_3$ [58, 59]. However, LaMnO$_3$ with cation deficiency or oxygen-excess tends to be a ferromagnetic metal with a mixture of Mn$^{3+}$ and Mn$^{4+}$ (*i.e.*, hole doping) [60], which is not consistent with the observation that our films are electron doped with nominal cation stoichiometry and insulating behavior [25]. Therefore, in our work, the emergence of electron doping at the LaMnO$_3$/SrTiO$_3$ interface triggered by the polar mismatch induces a mixed valence (*i.e.*, Mn$^{2+}$/Mn$^{3+}$) which is thought to give rise to the ferromagnetic phase [52, 54-57]. In fact, both the XAS and XMCD spectra observed in our relatively thick LaMnO$_3$/SrTiO$_3$ heterostructures (Fig. 1a,c) are similar to that of electron-doped manganites deriving their effects from chemical alloying (*e.g.*, Ce-doped LaMnO$_3$), which also exhibit an insulating ferromagnetic ground state [57].

The absence of ferromagnetism for films less than 5 UC is likely due to "over accumulation" of Mn$^{2+}$ (stemming from the need to localize a significant number of electrons to accommodate the potential and resulting in reduced magnetization from a lack of double-exchange coupling with $Mn^{3+}$) which favors an antiferromagnetic state [55]. In turn, we explored the possibility of reducing the $Mn^{2+}$ component by adjusting the composition of the films to confirm our understanding of the origin of the ferromagnetism and demonstrating the potential of such systems. Previous studies have found that La-vacancies give rise to hole doping in $LaMnO_3$ [58, 60]. Thus, we have grown 5% La-deficient $LaMnO_3$ ($La_{0.95}MnO_3$) films of varying thicknesses on $SrTiO_3$ substrates [25] to induce holes which should act to counter-dope the system under the influence of the potential and thus reduce the amount of $Mn^{2+}$ in the ultra-thin films (*i.e.*, drive it back from an over-electron-doped state to a more "optimally-doped" level). La-deficiency is an effective route to introduce holes into $LaMnO_3$ via the defect formula $(La^{3+}_{1-x}[V^{3-}_{La}]_x)(Mn^{3+}_{1-3x}Mn^{4+}_{3x})O^{2-}_3$. The $La_{0.95}MnO_3$ exhibits alternating $(La_{1-x}O)^{1-3x}$ and $(Mn^{3+}_{1-3x}Mn^{4+}_{3x}O_2)^{-1+3x}$ layers and thus a polar mismatch still exists at the $La_{0.95}MnO_3/SrTiO_3$ interface. Due to the La-deficiency, the magnitude of the polar discontinuity would be reduced with additional holes. From XAS studies of these $La_{0.95}MnO_3$ films, a weaker thickness dependence of the Mn valence is observed (Fig. 3a). In particular, the peak corresponding to $Mn^{2+}$ at 639.8 eV is suppressed in the 3 UC thick $La_{0.95}MnO_3$ films. In turn, subsequent XMCD studies reveal clear dichroism even in 3 UC thick $La_{0.95}MnO_3/SrTiO_3$ heterostructures (Fig. 3b) – confirming the importance of controlling the valence state of the material to relieve the over accumulation of electrons. All told, the observation of an insulating ferromagnetic ground state in films as thin as just 3 UC shows the power of emergent phenomena and approaches the limits of our ability to synthesize and study emergent phenomena at interfaces, and has potential applications in spin polarized tunneling devices.

To summarize, we have provided direct evidence for electron accumulation and ferromagnetism occurring within the polar, antiferromagnetic insulator $LaMnO_3$ when grown on non-polar SrTiO$_3$. Using XAS combined with first-principles calculations, the critical thickness for the onset of electron accumulation is determined to be 2 UC. The strength of the polar mismatch can drive "over-doping" of the LaMnO$_3$ which suppresses the onset of ferromagnetism as the average valence state tips towards Mn$^{2+}$. In stoichiometric LaMnO$_3$, ferromagnetism is observed in only 5 UC thick films. In turn, through chemical doping (achieved via control of the film stoichiometry), the average valence state can be tuned, and clear ferromagnetism can be observed in La$_{0.95}$MnO$_3$ films as thin as 3 UC. Ultimately, this work demonstrates the state-of-the-art as it pertains to ultra-fine control of materials, whereby controlling both the atomic-structure of interfaces and doping level, unprecedented properties and control of materials is produced.

## Acknowledgement

Z.H.C. would like to thank Dr. Chang-Yang Kuo for useful discussion and acknowledge support from the Laboratory Directed Research and Development Program of Lawrence Berkeley National Laboratory under U.S. Department of Energy Contract No. DE-AC02-05CH11231. Z.C. and L.-W.W. acknowledges the support from the U.S. Department of Energy, Office of Science, Office of Basic Energy Sciences, Materials Sciences and Engineering Division, of the under Contract No. DE-AC02-05-CH11231 within the Non-Equilibrium Magnetic Materials program (MSMAG). Z.Q.L. acknowledges financial support of the startup grant from Beihang University, China. X.R.W. acknowledges supports from startup grant from Nanyang Technological University. L.R.D. acknowledges support from the Department of Energy under Grant No. DE-SC0012375. R.X. acknowledges support from the National Science Foundation under grant DMR-1608938. R.G. acknowledges support from the National Science Foundation under grant OISE-1545907. T.V. acknowledges support from the Singapore National Research Foundation (NRF) Competitive Research Program (CRP Award No. NFR-CRP13-2014-04). J.L. acknowledges support from the Science Alliance Joint Directed Research and Development Program at the University of Tennessee. L.W.M. acknowledges supports from the Gordon and Betty Moore Foundation's EPiQS Initiative, Grant GBMF5307. Work performed at the electron microscopy facility of the Cornell Center for Materials Research was funded by the National Science Foundation (NSF) Materials Research Science and Engineering Centers program (DMR 1120296) and U.S. Department of Energy, Office of Basic Energy Sciences, Division of Materials Sciences and Engineering, under Award No. DE-SC0002334. The Advanced Light Source is supported by the Director, Office of Science, Office of Basic Energy Sciences, of the US DOE under Contract DE-AC02-05CH11231.

## Author Contributions

Z.H.C. performed the XAS, XLD, and XMCD measurements and interpreted the data with the assistance of J.L., A.T.N., and E.A.. C.J.L, X.W., L.S.F., and V.T. contributed materials. M.M., W.M.L., and L.R.D. contributed the RBS and PIXE data. M.E.H. performed the STEM and EELS experiments and interpreted the data with the help of J.A.M. and D.M. Z.H.C. and Z.Q.L. performed the X-ray diffraction, atomic force microscopy, SQUID, and transport studies. Z.C. and L.-W.W. provided the theoretical support. The manuscript was prepared by Z.H.C. and L.W.M.. The project was designed by Z.H.C., Z.Q.L., and J.L.. All authors read and contributed to the manuscript and the interpretation of the data.

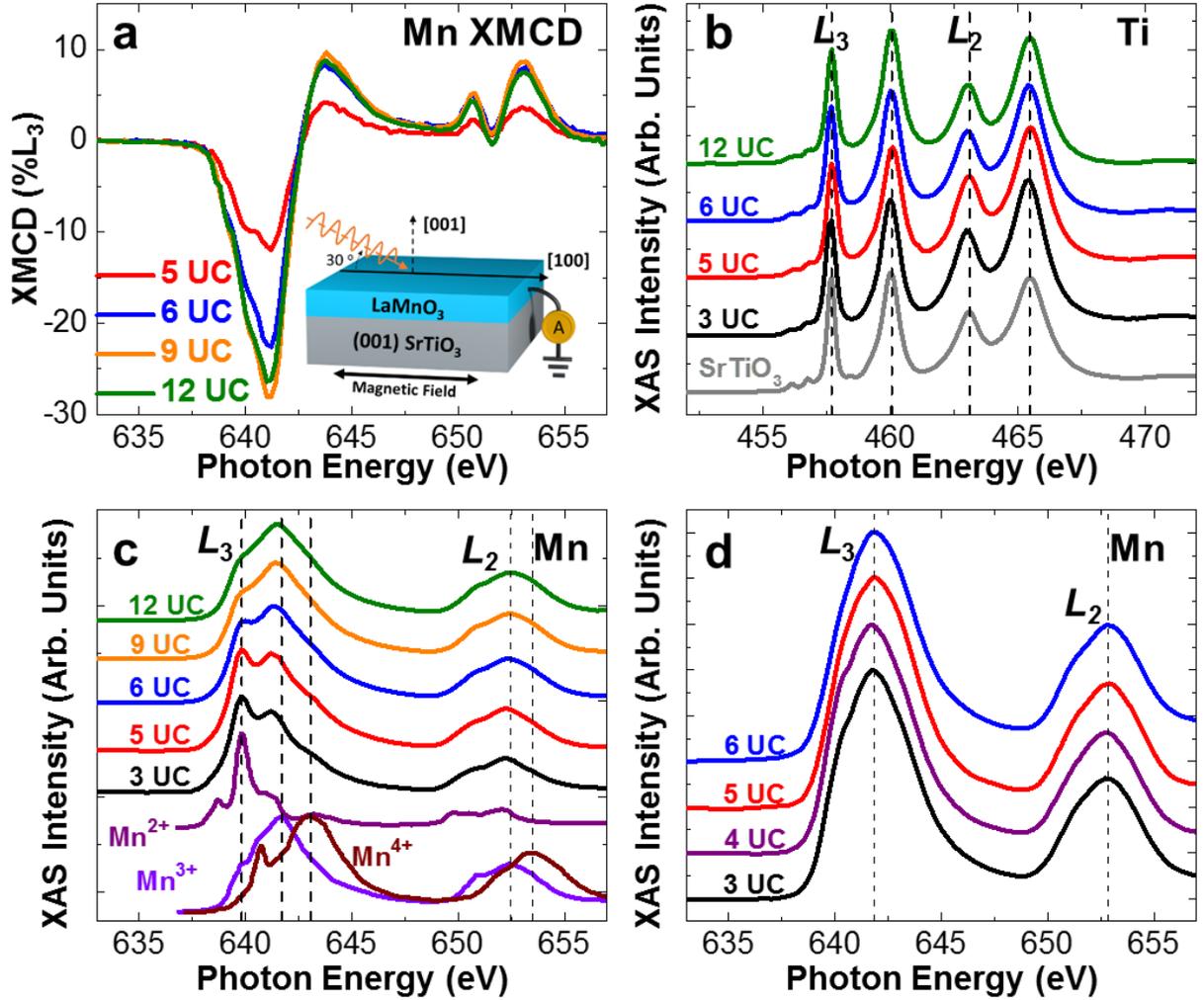

FIG. 1. (a) Mn XMCD spectra for various LaMnO$_3$/SrTiO$_3$ heterostructures. The inset shows a schematic of the experimental configurations for the X-ray spectroscopy studies. Thickness dependence of the XAS spectra of the LaMnO$_3$/SrTiO$_3$ heterostructures at the (b) Ti $L_{2,3}$, and (c) Mn $L_{2,3}$ edges with reference spectra for SrTiO$_3$, bulk SrMnO$_3$, LaMnO$_3$, and MnO for comparison. (d) XAS of Mn $L_{2,3}$ edges for various LaMnO$_3$/NdGaO$_3$ heterostructures where no reconstruction has occurred.

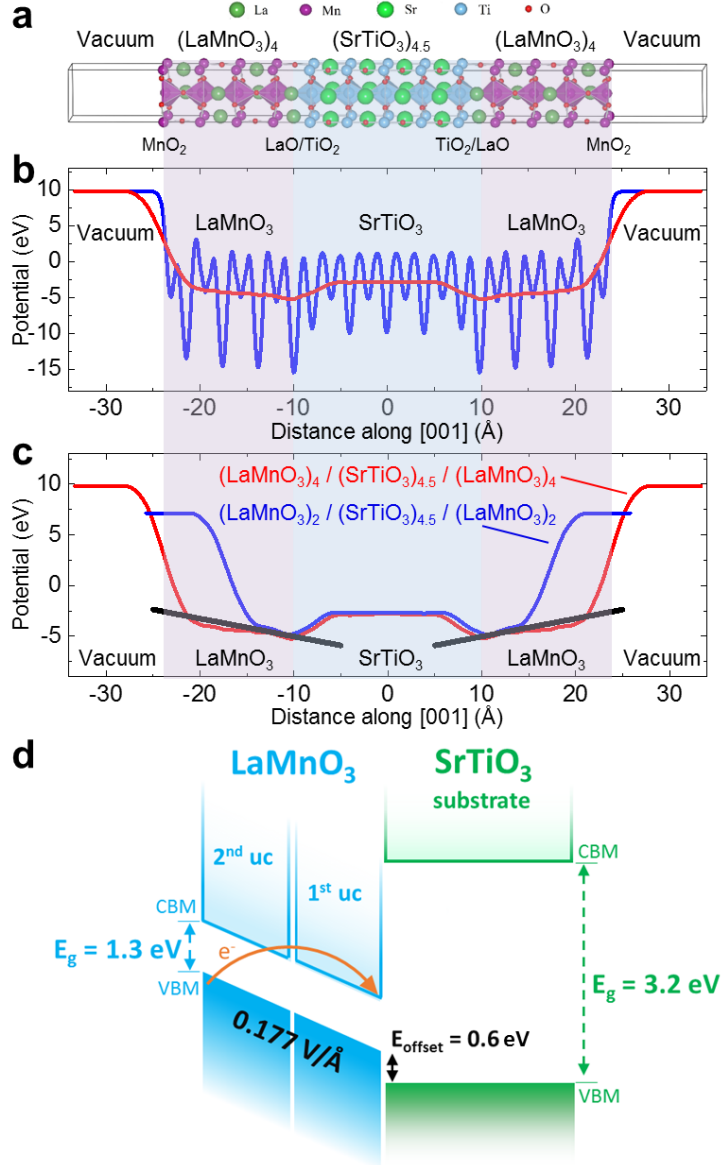

FIG. 2. (a) Schematic of the symmetric, vacuum-terminated $(LaMnO_3)_m/(SrTiO_3)_n/(LaMnO_3)_m$ slab with two identical *n*-type interfaces (here m = 4 and n = 4.5). (b) The in-plane average (oscillating blue line) and macroscopic average (red line) electrostatic potential across the $(LaMnO_3)_4/(SrTiO_3)_{4.5}/(LaMnO_3)_4$ simulation slab. (c) Comparison of the macroscopic average potential between the $(LaMnO_3)_4/(SrTiO_3)_{4.5}/(LaMnO_3)_4$ and $(LaMnO_3)_2/(SrTiO_3)_{4.5}/(LaMnO_3)_2$ heterostructures; the average intrinsic electric field is indicated by the black lines with a value of 0.177 V/Å. (d) Schematic band diagram of $LaMnO_3/SrTiO_3$ interface.

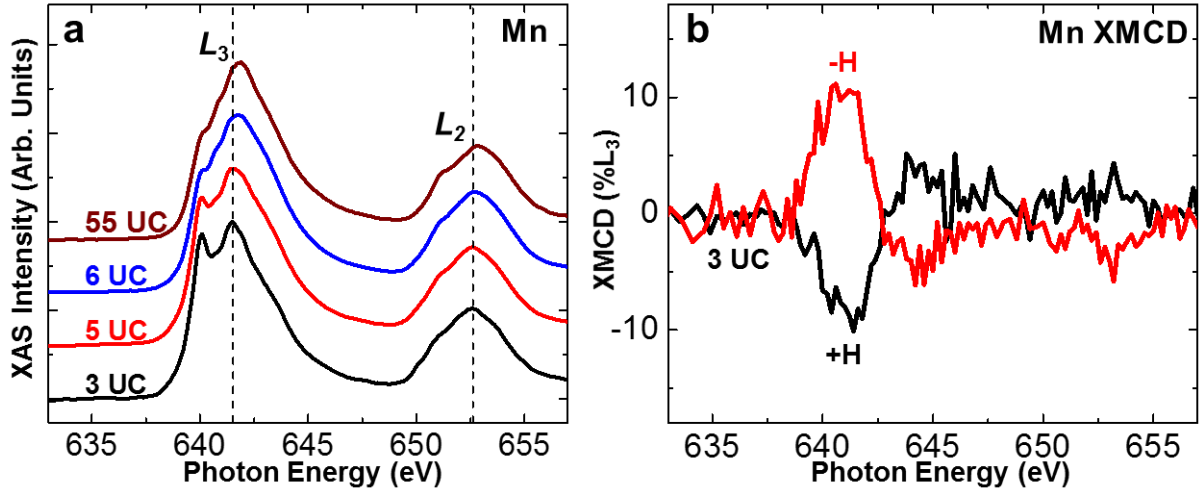

FIG. 3. (a) Mn $L_{2,3}$ XAS spectra of the $La_{0.95}MnO_3/SrTiO_3$ heterostructures. (b) Mn XMCD spectra for the 3 UC $La_{0.95}MnO_3/SrTiO_3$ heterostructure. The XMCD features reverse in sign when the magnetic field is reversed, confirming the reliability of the observation.

# Electronic Accumulation and Emergent Magnetism in LaMnO$_3$/SrTiO$_3$ Heterostructures

Zuhuang Chen[1,2], Zhanghui Chen[2], Z. Q. Liu[3], M. E. Holtz[4], C. J. Li[5,6], X. Renshaw Wang[7], W. M. Lü[8], M. Motapothula[6], L. S. Fan[9], J. A. Turcaud,[1] L. R. Dedon,[1] C. Frederick[10], R. J. Xu,[1] R. Gao,[1] A. T. N'Diaye[11], E. Arenholz[11], J. A. Mundy[1,2], T. Venkatesan[5,6], D. A. Muller[4], L.-W. Wang[2], J. Liu[10], and L. W. Martin[1,2]

[1] Department of Materials Science and Engineering, University of California, Berkeley, California 94720, USA

[2] Materials Science Division, Lawrence Berkeley National Laboratory, Berkeley, California 94720, USA

[3] School of Materials Science and Engineering, Beihang University, Beijing 100191, China

[4] School of Applied and Engineering Physics, Cornell University, Ithaca, New York 14853, USA

[5] Department of Materials Science and Engineering, National University of Singapore, Singapore 117575, Singapore

[6] NUSNNI-Nanocore, National University of Singapore, Singapore 117411, Singapore

[7] School of Physical and Mathematical Sciences & School of Electrical and Electronic Engineering, Nanyang Technological University, Singapore 637371

[8] Condensed Matter Science and Technology Institute, School of Science, Harbin Institute of Technology, Harbin 150081, People's Republic of China

[9] Center for Nanophase Materials Sciences, Oak Ridge National Laboratory, Oak Ridge, Tennessee 37831, USA

[10] Department of Physics and Astronomy, University of Tennessee, Knoxville, Tennessee 37996, USA

[11] Advanced Light Source, Lawrence Berkeley National Laboratory, Berkeley, California 94720, USA

## I. Sample fabrication and compositional analysis

LaMnO$_3$ films were grown on TiO$_2$-terminated SrTiO$_3$ (001) and GaO$_2$-terminated NdGaO$_3$ (110)$_o$ single-crystal substrates by reflection high-energy electron diffraction (RHEED)-assisted pulsed-laser deposition (Fig. S1). In this work, two sets of LaMnO$_3$/SrTiO$_3$ heterostructures were fabricated at two different laboratories. The first set of samples were grown at the National University of Singapore (NUS) as described elsewhere [1], and were determined to be stoichiometric (within 1-2%) via Rutherford backscattering spectrometry (RBS) measurements and particle-induced X-ray emission (PIXE) (Fig. S2a,b). The other set of samples were fabricated at Oak Ridge National Laboratory (ORNL), which were optimized to have 5% La deficiency as determined via RBS (Fig. S2c). All the films were deposited at 750°C, in a dynamic oxygen pressure of 7.5 mTorr, and a laser fluence of ~1.8 J/cm$^2$ from a stoichiometric, ceramic LaMnO$_3$ target. During the growth, clear RHEED intensity oscillations can be observed (Fig. S1b), indicating a layer-by-layer growth mode. Atomic force microscopy revealed that the film surfaces are atomically flat (Inset of Fig. S1b). In the samples produced at ORNL, these same conditions resulted in slightly La-deficient films (Fig. S2c). For the NUS samples, the target-to-substrate distance was ~80 mm and the laser spot size is about 5 mm$^2$; for the ORNL samples, the target-to-substrate distance was ~40 mm and the laser spot size is about 1.3 mm$^2$. After deposition, the samples were cooled down to room temperature under the deposition oxygen pressure.

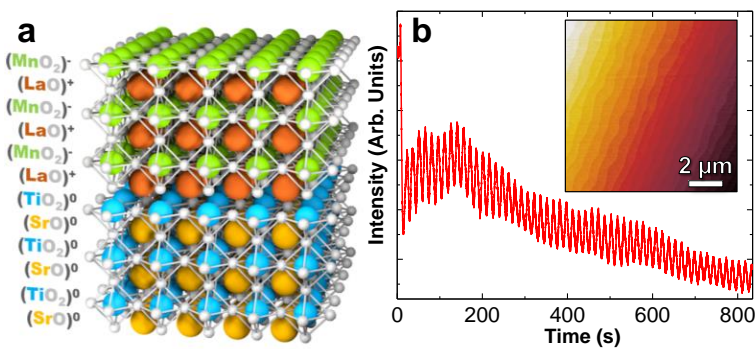

**Figure S1**. Atomically-engineered LaMnO$_3$/SrTiO$_3$ hetero-structures. (a) Schematic of LaMnO$_3$ grown on TiO$_2$-terminated SrTiO$_3$ (001) substrate. (b) Typical RHEED-intensity oscillations for the growth of a 55 UC LaMnO$_3$/SrTiO$_3$ (001) heterostructure; inset shows an AFM topography image of the same film revealing an atomically-smooth, terraced surface.

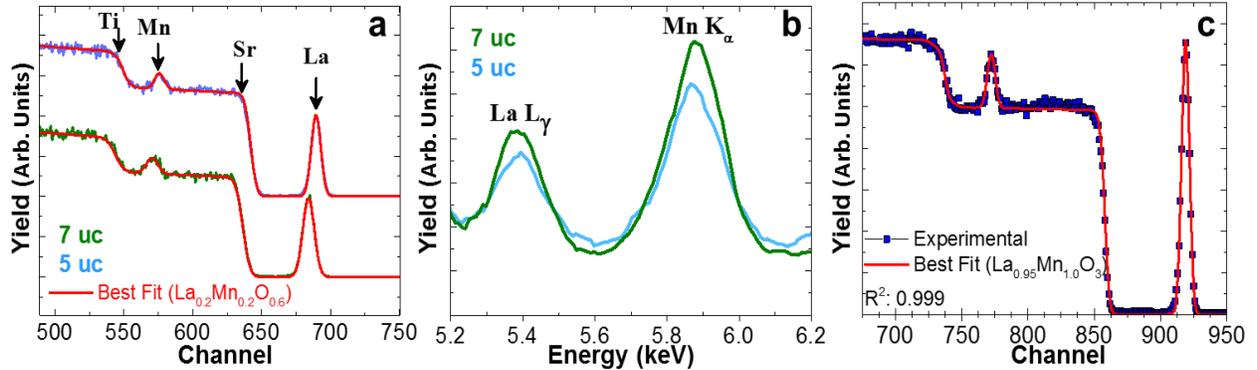

**Figure S2**. (a) RBS and (b) PIXE results for the LaMnO$_3$ heterostructures. Best fits of the data reveal a stoichiometric composition (within 1-2% error). (c) RBS results for the La$_{0.95}$MnO$_3$ heterostructures. Best fits of the data reveal a ~5% La-deficiency for the films.

The composition of the LaMnO$_3$ ultrathin films was characterized by a combination of RBS and PIXE using 2 MeV alpha particles at the Center for Ion Beam Applications (CIBA), Singapore. RBS spectra for LaMnO$_3$ (Fig. S2a) were collected using a Si surface barrier detector with a resolution of 19 keV full-width-at-half-maximum for multiple incident angles near 80° while the scattering angle was fixed at 160°. For all RBS studies, the fitting was done by SIMNRA software[9] with an uncertainty estimated to be 1-2%. PIXE spectra (Fig. S2b) were collected using a Si (Li) detector keeping the same distance for all measurements, and the La/Mn ratio was derived from the area ratio between the La $L_\gamma$ and Mn $K_\alpha$ peaks. RBS for the La-deficient heterostructures (Fig. S2c) was measured using 3.04 MeV alpha particles with incident angle $\alpha = 22.5°$, exit angle $\beta = 25.35°$, and scattering angle of 168° at Lawrence Berkeley National Laboratory.

## II. Structural characterization using X-ray diffraction

Single-phase, epitaxial growth of LaMnO$_3$/SrTiO$_3$ (001) heterostructures was confirmed by $\theta - 2\theta$ X-ray diffraction studies (Fig. S3) where only 00$l_{pc}$-diffraction peaks from the substrate and LaMnO$_3$ film are observed. Thickness fringes, apparent near the LaMnO$_3$ diffraction peaks, indicate the high crystalline quality of the films. At room temperature, bulk LaMnO$_3$ exhibits an orthorhombic structure with $a_o = 5.539$ Å, $b_o = 5.699$ Å, and $c_o = 7.718$ Å [2], corresponding to a pseudocubic unit cell with lattice constant $a_{pc} = 3.935$ Å. Therefore, epitaxial growth of LaMnO$_3$ on SrTiO$_3$ (cubic, $a = 3.905$ Å) leads to compressively strained films, which are confirmed by the expansion of the out-of-plane lattice parameter ($c = 3.94$ Å) in the X-ray diffraction patterns.

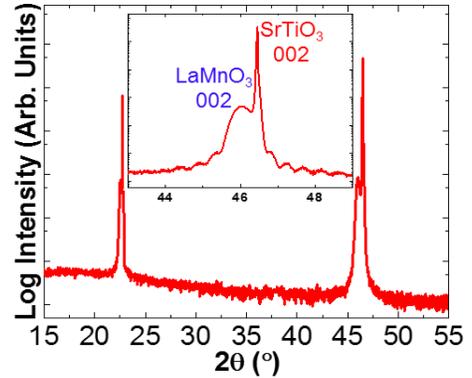

**Figure S3**. X-ray diffraction $\theta - 2\theta$ scan of a 55 unit cell thick LaMnO$_3$ film grown on a SrTiO$_3$ (001) substrate.

## III. STEM-EELS analysis

High-angle annular dark-field (HAADF) scanning transmission electron microscopy (STEM) and electron energy-loss spectroscopy (EELS) studies were carried out to characterize the film structure and interface abruptness. Cross-sectional TEM specimens were prepared by standard focused ion beam lift out techniques. STEM and EELS data were performed on the specimens in a 100 keV NION UltraSTEM; a 5th-order aberration-corrected microscope. The NION UltraSTEM was optimized for EELS spectroscopic imaging with a probe size of ~1Å, an EELS energy resolution of 0.75 eV, and a beam current of 100–200 pA. Spectroscopic maps of the La $M_{4,5}$ edges, the Mn $L_{2,3}$ edges, and the Ti $L_{2,3}$ edges were acquired with an energy dispersion of 0.25 eV/channel with a Gatan Quefina dual-EELS spectrometer with a 2 ms dwell time per scan position. The EELS edges were integrated after background subtraction, fitted by a linear combination of power laws.

HAADF-STEM images of a representative 12 UC thick LaMnO$_3$/SrTiO$_3$ (001) heterostructure reveals that the heterointerface is coherent without misfit dislocations (Fig. S4). Note that the top two unit cells of the film are amorphous due to electron beam damage during acquisition of the spectroscopic image. Further analysis using electron energy loss spectroscopy (EELS) provides both elemental maps and concentration line profiles across the interface for the heterostructure (Fig. S5). The HAADF-STEM imaging and the STEM-EELS suggest a high-quality film with a sharp interface displaying only 2-3 UC of interdiffusion. An asymmetry in cation intermixing is observed wherein the *B*-site cations (*i.e.*, Mn and Ti) show an atomically-sharp interface (only one unit cell of interdiffusion) and the *A*-site cations (*i.e.*, La and Sr) show considerable interdiffusion (at least 2-3 unit cells). We find the interfaces in the LaMnO$_3$/SrTiO$_3$ system studied herein are of comparable quality of those investigated in previous studies in related systems [3,4], suggesting that intrinsic physical phenomena can be measured from these heterostructures.

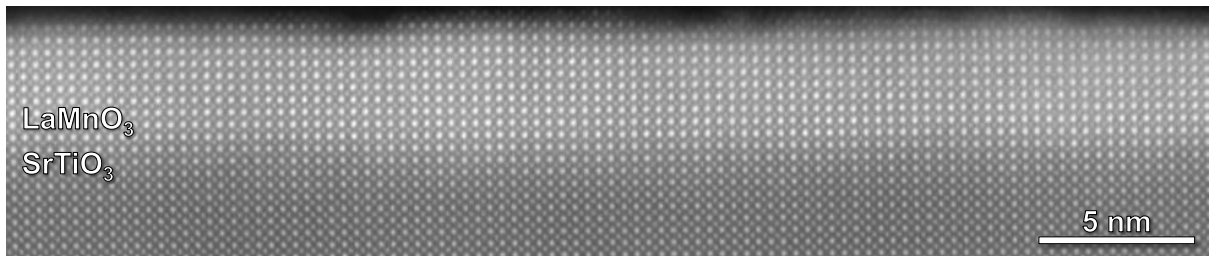

**Figure S4**. HAADF-STEM image of a 12 UC LaMnO$_3$/SrTiO$_3$ (001) heterostructure.

Although there is slight La interdiffusion, previous studies have suggested that *A*-site interdiffusion in related structures would not preclude polarization build-up and eventual electronic reconstruction, but that preferential *A*-site interdiffusion might actually enhance the potential build-up [3]. And the slight interdiffusion observed above cannot account for the large amount of $Mn^{2+}$ observed in the ultrathin films. Moreover, either diffusion of a Mn atom onto a Ti site or diffusion of a Sr atom onto a La site should drive the Mn valence toward +4 [3]. Therefore, the changes in manganese valence (towards $Mn^{2+}$ as film thickness reduces) observed from XAS in Fig. 1c cannot be interpreted by interdiffusion-driven valence changes.

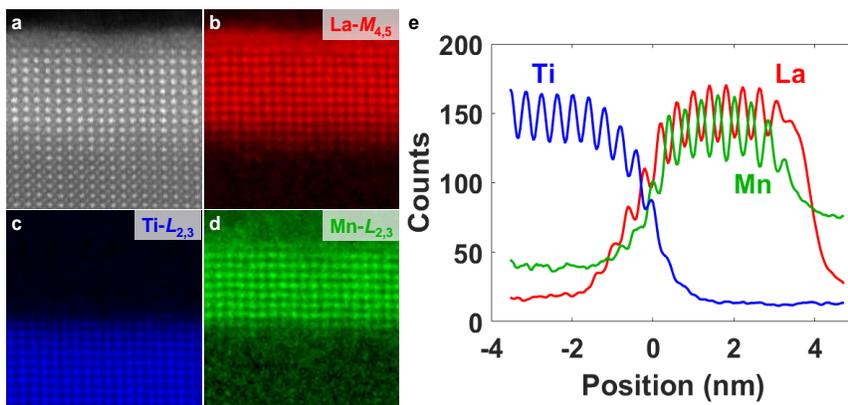

**Figure S5.** Atomic resolution EELS images of the LaMnO$_3$ (12 UC) /SrTiO$_3$ heterostructure. (a) HAADF image near the interface. (b-d) EELS elemental maps of La $M_{4,5}$, Ti $L_{2,3}$, and Mn $L_{2,3}$ edges, respectively. (e) Line profiles of La $M_{4,5}$, Ti $L_{2,3}$, and Mn $L_{2,3}$ edges across the interface.

IV.     **Soft X-ray spectroscopy measurements**

X-ray spectroscopy measurements were carried out at beamlines 4.0.2 and 6.3.1 of the Advanced Light Source, Lawrence Berkeley National Laboratory. The X-ray absorption spectroscopy (XAS) measurements were performed at room temperature with fixed circularly-polarized light without applied magnetic field using total-electron-yield (TEY) mode and an X-ray angle of incidence of 30° to the sample surface (so-called grazing incidence geometry). The intensity of the incident X-ray beam and the sample drain current were simultaneously recorded to normalize the obtained spectrum to the beam intensity. X-ray magnetic circular dichroism (XMCD) spectra were measured using fixed X-ray circular polarization and point-by-point reversal of the external magnetic field of magnitude 1.9 T along the X-ray beam direction [100] at 25 K, using TEY detection also in grazing incidence geometry. The dichroism is the difference between the two spectra measured with parallel and anti-parallel alignment of the magnetization direction with the photon helicity vector. To ensure that the XMCD signal is of magnetic origin, we repeated the measurement with opposite light polarization (left or right circular polarized light) and confirmed that the sign of the XMCD reverses. X-ray linear dichroism (XLD) measurements were performed at room temperature and were obtained from the difference of horizontal and vertical polarized light absorption spectra ($I_h - I_v$) without applied magnetic field in the TEY configuration. Since the magnetic ordering temperature of LaMnO$_3$ thin films is well below room temperature, the XLD signal is solely determined by the orbital anisotropy. The X-ray beam was incident on the sample at an angle of 20° from the sample surface; the light polarizations were selected by using an elliptically polarized undulator. In such a grazing measurement geometry the X-ray electric field is oriented parallel to the surface for vertically polarized light, and almost perpendicular to the surface for horizontally polarized light, thus providing maximum sensitivity to changes in orbital character along the different crystallographic axes. Spectra were captured with the order of polarization rotation reversed (*e.g.*, horizontal, vertical, and then vertical, horizontal) so as to eliminate experimental artifacts.

To determine the orbital occupancy and its evolution with film thickness, we performed XLD measurements at the Mn $L_{2,3}$ edges on various thickness LaMnO$_3$/SrTiO$_3$ heterostructures in grazing incidence geometry (Fig. S6a). In this geometry, the X-ray electric field is oriented parallel to the surface (along in-plane [010]) for vertically-polarized light and almost perpendicular to the surface (along out-of-plane [001]) for horizontally-polarized light. The XLD is calculated as the intensity difference ($I_h - I_v$) between the XAS spectra measured with horizontal ($E_h$) and vertical ($E_v$) linear polarized light. For rare-earth manganites, the XLD at the Mn $L_{2,3}$ edges gives information on the empty *d* orbital states wherein larger (smaller) absorption intensity for in-plane polarized light corresponds to more in-plane (out-of-plane) empty states in the $e_g$ band and thus a preferential occupation of out-of-plane $d_{3z^2-r^2}$ (in-plane $d_{x^2-y^2}$) orbitals [5,6]. Similar Mn $L_{2,3}$ XLD features (Fig. S6b) are observed in the LaMnO$_3$/SrTiO$_3$ heterostructures with differing LaMnO$_3$ thicknesses wherein negative XLD signals are observed, indicating a preferential occupation of the out-of-plane $d_{3z^2-r^2}$ orbitals. Such symmetry breaking is due to the epitaxial strain from the underlying substrate [5]. The same XLD feature observed in various LaMnO$_3$ film thicknesses indicates that the magnetic phase transition observed in LaMnO$_3$/SrTiO$_3$ heterostructures is not driven by orbital reconstruction.

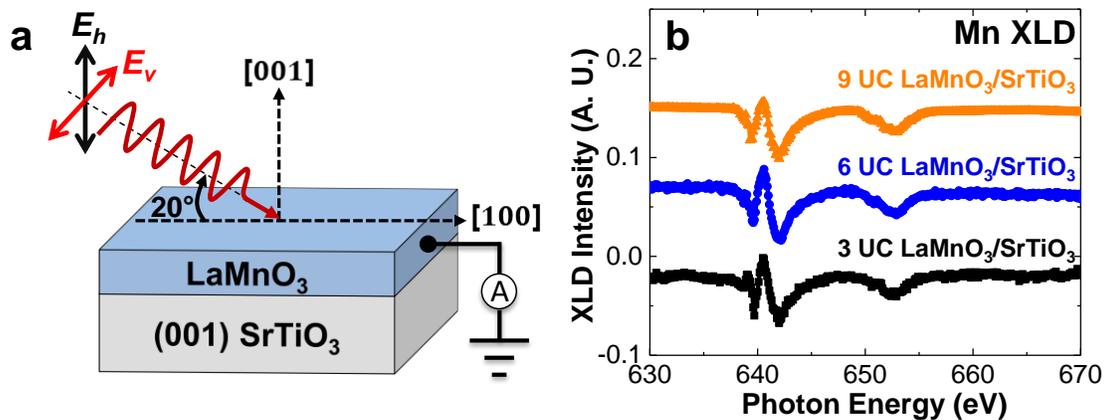

**Figure S6**. (a) A sketch of the XLD measurements. (b) XLD spectra acquired at the Mn $L_{2,3}$ edges at 300 K for 3 UC LaMnO$_3$/SrTiO$_3$ (black data, bottom), 6 UC LaMnO$_3$/SrTiO$_3$ (blue data, middle), and 9 UC LaMnO$_3$/SrTiO$_3$ (orange data, top) heterostructures.

## V. XAS of 2 UC LaMnO$_3$/SrTiO$_3$ heterostructures

The XAS spectrum about the Mn $L_{2,3}$ edges for a 2 UC thick LaMnO$_3$/SrTiO$_3$ heterostructure reveals a predominant absorption peak at ~639.8 eV corresponding to Mn$^{2+}$ (Fig. S7). The appearance of the strong signature of Mn$^{2+}$ in the 2 UC thick sample suggests that electronic reconstruction has already occurred, consistent with the theoretical calculations.

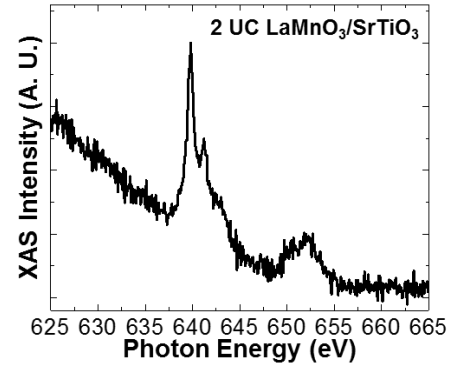

**Figure S7**. XAS spectrum about the Mn $L_{2,3}$ edges for a 2 unit-cell-thick LaMnO$_3$/SrTiO$_3$ heterostructure.

## VI. XAS about the O *K* edge of the LaMnO$_3$/SrTiO$_3$ heterostructures

The proposed electronic reconstruction in the LaMnO$_3$/SrTiO$_3$ heterostructures is further supported by thickness-dependent XAS spectra about the O *K* edge (Fig. S8). It should be noted that the features of the spectrum from a bare SrTiO$_3$ substrate have significant contributions to those spectra from 3 UC and remain visible in the spectra of the 5 and 6 UC thick samples. This provides a rough probe of the depth-resolution of the surface-sensitive XAS studies. For the LaMnO$_3$, the O $K$ edge XAS probes the unoccupied density of states with O 2$p$ symmetry due to dipole selection rules that arise mainly from the hybridization of O 2$p$ states with Mn 3$d$ ($hv$ = 525 - 529 eV) and La 4$f$/5$d$ ($hv$ = 529 -536 eV) states [7]. The region above 536 eV is attributed to mixing with higher energy states of the Mn and La atoms, such as Mn 4$sp$ and La 6$sp$ [8]. Therefore, the relevant peak for the study of Mn valence change is the first peak at around 527-528 eV, which is due to dipole transitions from O 1$s$ to O 2$p$ states, which are hybridized with the unoccupied Mn 3$d$ orbitals. The intensity of this peak represents the 2$p$ hole and is also an indirect measure of the Mn 3$d$ level occupancy. As shown, the first peak gradually shifts to a lower photon energy with the increasing LaMnO$_3$ thickness, indicating an increase of the oxidation state of the Mn atom [8,9].

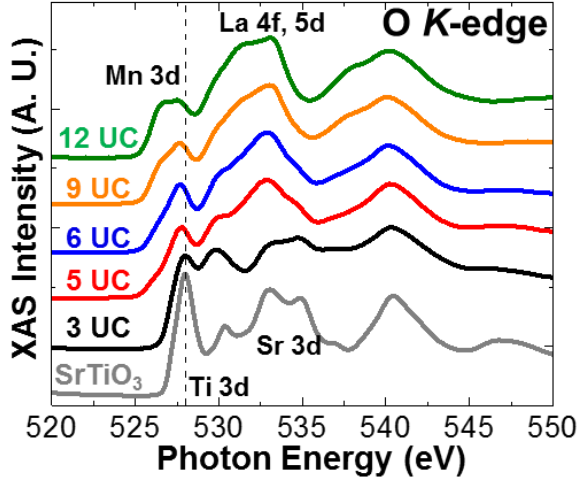

**Figure S8**. Thickness dependence of the XAS spectra of the O $K$ edge of LaMnO$_3$/SrTiO$_3$ heterostructures with that for a bare SrTiO$_3$ substrate provided for comparison.

## VII. First-principles calculations

First-principles calculations are performed using the projector augmented wave (PAW) [10,11] formalism and a plane wave basis set, as implemented in the Vienna *ab initio* simulation package (VASP) [12,13]. The exchange and correlation potential are treated in the framework of generalized gradient approximation (GGA) of Perdew-Burke-Ernzerbof (PBE) [14]. Local spin-density approximation with an additional Hubbard interaction (LSDA+U) was used for the exchange-correlation functional. The effective Hubbard parameter U$_{eff}$ is set to 8.5 eV for $d$ electrons of Sr and Ti atoms, and 3.5 for $d$ electrons of Mn atoms. We found these values can well reproduce the experimental band gap of bulk SrTiO$_3$ (3.2 eV) and LaMnO$_3$ (1.3 eV) (Figure S9a,b). All of the calculations consider spin polarization and the LaMnO$_3$ part of the slab is A-type antiferromagnetic ordered. For the summation of charge densities over the

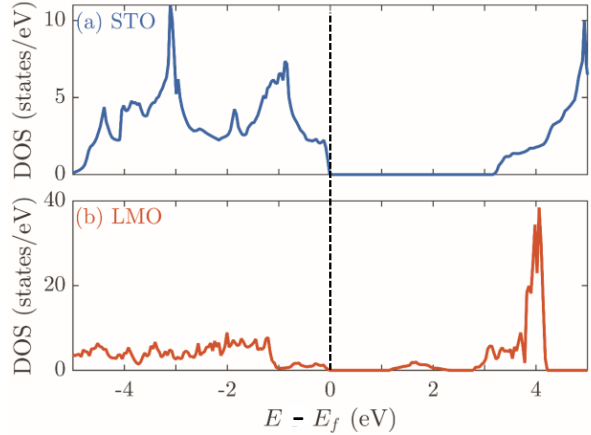

**Figure S9**. Total density of states (DOS) (states/eV per LaMnO$_3$ or SrTiO$_3$ molecule) of (a) bulk SrTiO$_3$, and (b) bulk LaMnO$_3$. The referenced zero energy (black dashed line) is set to the Fermi level. Bulk LaMnO$_3$ was calculated in an A-type antiferromagnetic ordered state with 20-atoms supercell.

Brillouin zone [15], an 8×8×8 $k$-point mesh for bulk and 5×5×1 $k$-point mesh for slab are adopted in the calculation of the total energy and ionic forces. The wave functions are expanded in plane waves up to a cutoff of 500 eV and the convergence precision of the total energy is set to be lower than 10$^{-5}$ eV. The atomic force is relaxed to be lower than 0.02 eV/Å. To evaluate the intrinsic electric field, we constructed a symmetric (LaMnO$_3$)$_m$(SrTiO$_3$)$_n$(LaMnO$_3$)$_m$ slab, where the LaMnO$_3$ film is MnO$_2$-terminated on the vacuum side and LaO-terminated on the SrTiO$_3$ side while the SrTiO$_3$ film is TiO$_2$-terminated on both sides. The value n for SrTiO$_3$ slab is set to 4.5 (*i.e.*, the layer size of SrTiO$_3$ part is $\sqrt{2}\times\sqrt{2}\times 4.5$). For LaMnO$_3$, we studied systems with the value $m = 2, 4, 6$. Slabs are separated by 20 Å of vacuum to eliminate the spurious slab-slab interaction. The slope for the electric field in Fig. 2c is computed by the mean value of the two slabs with $m = 2, 4$. Note that there have already been charge transfer in the slab with $m = 2$. Further results for models of a (LaMnO$_3$)$_6$/(SrTiO$_3$)$_{4.5}$/(LaMnO$_3$)$_6$ heterostructure are provided (Figure S10). The calculations find a similar potential build-up in the LaMnO$_3$ layer with a slope of ~0.10 V/Å; smaller than that found in (LaMnO$_3$)$_m$/(SrTiO$_3$)$_{4.5}$/(LaMnO$_3$)$_m$ heterostructures (where $m = 2, 4$).

This is reasonable since the transferred charge will partially balance the internal field, resulting in a decrease of the observed slope in thicker slabs, and eventually the potential field will disappear in the bulk. Such a thickness dependence of the potential slope is similar to what was found in LaAlO$_3$/SrTiO$_3$ [16-18]. Thus, the actual screened-field value will be larger than the observed 0.177 V/Å.

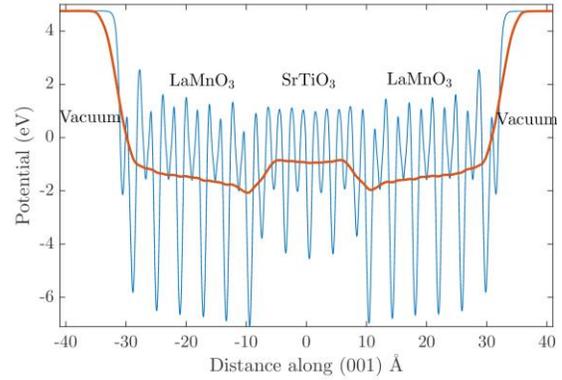

**Figure S10.** The in-plane average (oscillating blue line) and its macroscopic average (red line) electrostatic potential across the (LaMnO$_3$)$_6$/(SrTiO$_3$)$_{4.5}$/(LaMnO$_3$)$_6$ simulation slab.

For the band alignment calculations, we first compute the difference between the valence band maximum (VBM) and the average potential in LaMnO$_3$ ($\Delta E_{LaMnO_3}$) and SrTiO$_3$ ($\Delta E_{SrTiO_3}$) bulk, respectively. Then, their difference is obtained: $\Delta E_{bulk} = \Delta E_{LaMnO_3} - \Delta E_{SrTiO_3}$. We further compute the difference ($\Delta E_{ave}$) between the average potential of LaMnO$_3$ and SrTiO$_3$ part in the modeled slabs. The final VBM offset is: $\Delta E_{VBM} = \Delta E_{bulk} + \Delta E_{ave}$.

Our DFT calculations indicate that electron accumulation has occurred in the LaMnO$_3$/SrTiO$_3$. For example, the number of valence band electrons of Mn cations is 11.540, 11.586, 11.610, and 11.638 as one transitions from the surface layer to the interface layer in the model of (LaMnO$_3$)$_4$/(SrTiO$_3$)$_{4.5}$/(LaMnO$_3$)$_4$. This slope is steeper in the model of (LaMnO$_3$)$_2$/(SrTiO$_3$)$_{4.5}$/(LaMnO$_3$)$_2$, and the corresponding number of valence band electrons of Mn cations is 11.550 and 11.615 when going from the surface layer to the interface layer.

Note that our above calculated potential build-up is based on an idealized and perfectly abrupt interface without surface defects and adsorbates. We further calculated the effect of oxygen vacancies at the film surface on the electrostatic potential of the LaMnO$_3$/SrTiO$_3$ heterostructures.

To do so, we eliminated one oxygen atom in the surface $MnO_2$ layer at both sides of the $(LaMnO_3)_4/(SrTiO_3)_{4.5}/(LaMnO_3)_4$ slab (Fig. 2a, main text). This results in one missing oxygen ions per 24 oxygen ion in each $LaMnO_3$ slab – corresponding to an oxygen vacancy of 4.2%. The in-plane- and macroscopically-averaged electrostatic potential across the slab was then calculated and plotted (Figure S11).

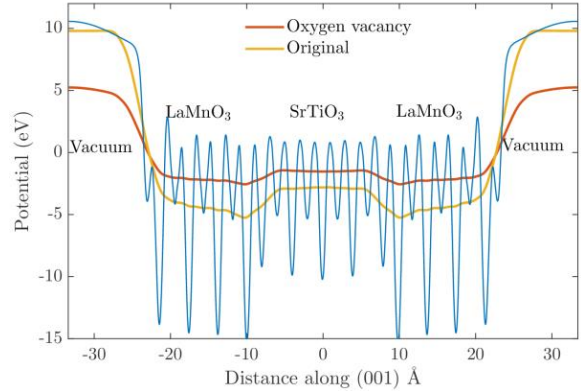

**Figure S11.** The in-plane average (oscillating blue line) and its macroscopic average (red line) electrostatic potential across the $(LaMnO_3)_4/(SrTiO_3)_{4.5}/(LaMnO_3)_4$ simulation slab in the case of surface oxygen vacancies. As a comparison, the yellow line indicates the macroscopic average for the original case without oxygen vacancies.

Overall, the built-in potential is changed when oxygen vacancies are present, for example, in the new cell with 4.2% oxygen vacancy concentration, the potential is reduced from 0.177 V/Å to 0.07 V/Å. The reason for this change, is that the charge from the oxygen vacancies partially compensates the band bending. However, having oxygen vacancies alone as the compensation mechanism is unlikely for two reasons: 1) even a large oxygen-vacancy concentration of 4.2% is insufficient to fully compensate the potential build-up. Note that our films were grown at a relatively high oxygen pressure of $10^{-2}$ Torr and thus the creation of such a high concentration of oxygen vacancies is highly unlikely. 2) We observe a critical thickness for the onset of electron accumulation of 2 unit cells. This is consistent with the theoretical thickness required for a sufficiently strong built-in potential which triggers charge transfer towards the interface in this model. But it is not consistent with having a small concentration of oxygen vacancies which would reduce the magnitude of the built-in potential, and thus increase the critical thickness and also not consistent with a large concentration of oxygen vacancies which would require no critical thickness for screening to occur. Ultimately, it is possible that the polar field (or built-in potential) can induce a number of potential screening mechanism including, formation of oxygen vacancies, cation intermixing, and lattice reconstruction. In that case, the excess electrons introduced by the oxygen vacancies in the film could, in turn, move to the interface to further compensate the potential built-in potential.

## VIII. SQUID magnetometry measurements

In-plane magnetization-magnetic field hysteresis loops (Fig. S12a) of a 9 UC thick $LaMnO_3/SrTiO_3$ heterostructure at 25 K (the same temperature as the XMCD data in the main text) reveal well-saturated hysteresis loops with a magnetization of ~400 emu/cc; indicating that the film is strongly ferromagnetic. Subsequent magnetization-temperature studies (Fig. S12b) done after cooling from room temperature under an applied field of 200 Oe and subsequently measured during heating at an applied field of 200 Oe reveal a potential magnetic Curie temperature for this heterostructure of ~120 K.

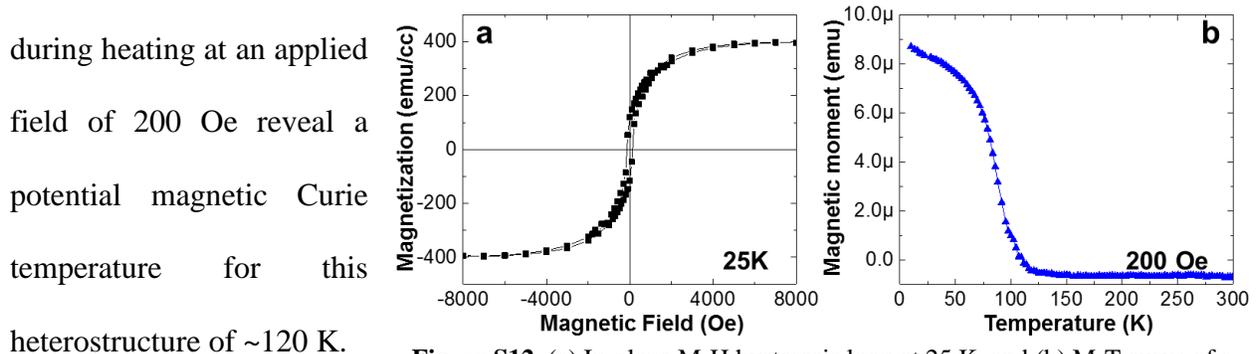

**Figure S12**. (a) In-plane M-H hysteresis loop at 25 K, and (b) M-T curve of a 9 UC $LaMnO_3/SrTiO_3$ heterostructure.

## IX. Transport measurements

Bulk $LaMnO_3$ is an A-type antiferromagnetic Mott insulator with bandgap of 1.3 eV. Previous studies have shown that off-stoichiometry would reduce the resistivity of the film and eventually turn it into ferromagnetic metal [19,20]. Collinear four probe resistance measurements (Fig. S13) using Physical Property Measurement System (PPMS, Quantum Design) found that all the films exhibit insulating behavior in a temperature range from 10 K to 350 K, consistent with previous study [1]. The high insulating nature indicates that our films are near-stoichiometric with low extrinsic growth-induced defects, consistent with our above composition analysis results using RBS and PIXE.

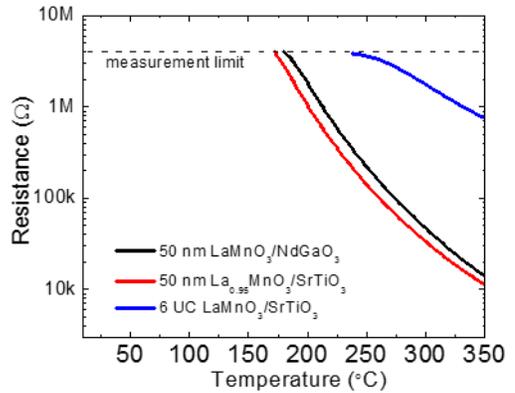

**Figure S13**. Resistance of the LaMnO$_3$ and La$_{0.95}$MnO$_3$ as a function of temperature.